\documentclass[twocolumn]{aa}
\usepackage{color}
\usepackage{graphicx}
\usepackage[usenames,dvipsnames]{xcolor}
\usepackage{natbib}
\usepackage{txfonts}
\usepackage[hyperfootnotes=true, hidelinks, colorlinks, citecolor=blue, linkcolor=blue, linktocpage, bookmarks=true]{hyperref}
\usepackage{footnote}
\usepackage{float}
\usepackage{multirow}
\usepackage{amsmath}

\usepackage[bottom, flushmargin]{footmisc}

\makeatletter
\renewcommand*\aa@pageof{, page \thepage{} of \pageref*{LastPage}}
\makeatother

\begin{document}

\title{Tidal perturbations and eclipse mapping in the pulsations in the hierarchical triple system U~Gru}
\titlerunning{Tidal perturbations and eclipse mapping in U~Gru}

\author{C. Johnston\inst{1,2}
    \and A. Tkachenko\inst{2}
    \and T. Van Reeth\inst{2}
    \and D. M. Bowman\inst{2}
    \and K. Pavlovski\inst{3}
    \and H. Sana\inst{2}
    \and S. Sekaran\inst{2}
    }

\institute{ Department of Astrophysics, IMAPP, Radboud University Nijmegen, P. O. Box 9010, 6500 GL Nijmegen, the Netherlands \\ \email{cole.johnston@ru.nl}
   \and Institute of Astronomy, KU Leuven, Celestijnenlaan 200D, 3001 Leuven, Belgium 
   \and Department of Physics, Faculty of Science, University of Zagreb, 10 000 Zagreb, Croatia
  }

\date{Received Date Month Year / Accepted Date Month Year}

\abstract {Unambiguous examples of the influence of tides on self-excited, free stellar pulsations have recently been observationally detected in space-based photometric data.} 
{We aim to investigate U~Gru and contextualise it within the growing class of tidally influenced pulsators. Initial analysis of U~Gru revealed  frequencies spaced by the orbital frequency that are difficult to explain by currently proposed tidal mechanisms.} 
{We re-investigate the TESS photometry of U~Gru alongside new {\sc uves} spectroscopy. We analyse the {\sc uves} spectroscopy with least-squares deconvolution and spectral disentangling techniques, and perform an atmospheric analysis. We remove the binary signature from the light curve using an effective model in order to investigate the pulsation signal in the residuals. We track the amplitudes and phases of the residual pulsations as a function of the orbital period to reveal their tidal influence.} 
{We establish that U~Gru is likely a hierarchical triple system. We identify a single p mode oscillation to exhibit amplitude and phase variation over the binary orbit. We propose a toy model to demonstrate that the series of frequencies separated by the orbital frequency can be reproduced by eclipse mapping. We find no evidence of modulation to the other independent oscillation modes.} 
{We demonstrate that U~Gru hosts at least one tidally perturbed pulsation. Additionally we argue that eclipse mapping of the dominant, tidally perturbed mode can produce the series of frequencies separated by the observed orbital frequency. Our simulations show that the effects of eclipse mapping are mode dependent, and are not expected to produce an observable signature for all pulsation modes in an eclipse binary. }

\keywords{ asteroseismology -- stars: binaries: eclipsing -- stars: individual: U~Gru -- stars: oscillations -- stars: binaries: close}
\maketitle


\section{Introduction}

Self-excited, free stellar pulsations give rise to surface flux variations that are observable as brightness variations. The presence of a close binary companion can cause modulations to these brightness variations as seen by the observer. The modulations can be induced by: (1) directly modifying the interior of the pulsating star giving rise to tidally perturbed pulsations \citep[e.g.,][]{Smeyers1983,Reyniers2003}; (2) altering the axis about which the pulsations occur \citep[e.g.,][]{Handler2020,Fuller2020}; or (3) through preferentially altering the visibility of the pulsating star's photosphere along the observer's line of sight \citep[e.g. eclipse mapping; ][]{Gamarova2003,Reed2005}. 

Recently, a new class of tidally tilted pulsators in binary systems has been observationally established wherein the tidal forces are assumed to be strong enough to re-align the pulsation axis with the line of apsides connecting the centres of mass of each component \citep{Fuller2020}. Since the pulsation axis is no longer aligned with the rotation axis in this scenario, a time-dependent observed amplitude modulation is introduced as the pulsation nodal lines periodically change visibility with orbital phase. This is comparable to the configuration observed in rapidly oscillating Ap stars in which the pulsation axis is inclined with respect to the rotation axis due to the strong magnetic fields of these stars, commonly referred to as the oblique pulsator model \citep{Kurtz1982,Bigot2002,Saio2004}. Furthermore, there is a subset of tidally tilted pulsators in which the tidal deformation of the pulsating star effectively traps the pulsation to a single side of the star due to mode coupling, i.e. tidally trapped pulsators \citep{Fuller2020}. This creates an apparent orbital phase dependence of the pulsation amplitude. When viewed in Fourier space, the amplitude modulation results in pulsation multiplets which have components separated by the orbital frequency. 

The phenomenon of tidally tilted pulsations has, so far, been observed in a handful of stars observed by NASA's Transiting Exoplanet Survey Satellite \citep[TESS; ][]{Ricker2015}. The cases of HD~74423, CO~Cam, and HD~265435 show pulsations that are preferentially trapped to one side of the star \citep{Handler2020, Kurtz2020,Jayaraman2022}. The example of TIC~63328020, however, shows non-axisymmetric sectoral modes which have a tilted pulsation axis and weak trapping \citep{Fuller2020,Rappaport2021}, while the oscillating eclipsing Algol-type \citep[oEA; ][]{Mkrtichian2003} systems TZ~Dra and HL~Dra show evidence for a tidally tilted pulsation axis \citep{Shi2021,Kahraman2022}. 

In addition to these clear examples of tidally tilted pulsators, there are a handful of objects that demonstrate other forms of tidal perturbations to the observed pulsations. Notably, the eclipsing binary RS~Cha shows multiple phenomena, including the tidal splitting of intrinsic pulsation modes induced by the tidal bulge acting as the axis of symmetry, as well as observed pulsation amplitude decrease due to the presence of eclipses \citep{Balona2018,Preece2019,Steindl2021}. Furthermore, there are now several (eclipsing) binaries pulsating in either pressure (p) or gravity (g) mode oscillations that show tidal perturbations or other distortions that modify the observed pulsation amplitude over the orbit \citep{Jerzykiewicz2020,Southworth2020,Southworth2021,Lee2021,VanReeth2022}. While the sample of tidally perturbed pulsators is increasing, there is no consensus on a proposed underlying mechanism. There is no obvious underlying commonality in orbital configuration, mass range, or previous evolutionary history amongst the known sample of tidally influenced pulsating stars \citep{Handler2022}. 

Given the recent increase in the identification of tidal signatures in a diversity of pulsating stars, astronomers are starting to utilise the high precision afforded by space-based telescopes to perform asteroseismology and probe the influence of tides on the stellar interior. However, to be able to exploit tidally influenced modes in asteroseismic analysis, we require the unambiguous identification of the underlying tidal phenomena in question. Therefore, complicated cases such as that of the oEA system U~Gru, which shows signatures of tidal effects as well as eclipse mapping \citep{Reed2005, Gamarova2003} in its pulsation modes \citep{Bowman2019}, need a detailed investigation of combined time-series photometry and spectroscopy before it can undergo a tidal asteroseismic analysis. 

U~Gru is a bright oEA system with a binary period of 1.8805~d, initially discovered to be an eclipsing binary with an A5 spectral type by \citet{Brancewicz1980}. Recently, U~Gru was observed by TESS in sectors 1 and 28, yielding 54 days of high precision space based photometry. Using the initial sector of TESS data, \citet{Bowman2019} report the discovery of several independent p mode pulsations in addition to a series of significant frequencies split by (multiples of) the orbital frequency. However, there is no clear symmetry as to how the frequencies that are split by the orbital frequency are distributed, making its interpretation as a tidal phenomenon challenging given the currently proposed mechanisms \citep[e.g.,][]{Balona2018,Fuller2020}. oEA systems are known to have recently undergone mass-transfer, effectively rejuvenating the accretor and placing it closer to the zero-age main-sequence (ZAMS). To further complicate matters, oEA systems are commonly observed to have distant tertiary companions, potentially introducing more complicated tidal forces in the system due to the ever changing tidal potential \citep{Fuller2013}. 

In this paper, we re-visit the TESS photometry of U~Gru alongside new {\sc uves} spectroscopy in order to identify the mechanisms causing the tidal effects observed in its pulsations. In Section~\ref{section:spectra} we present new spectroscopic observations and atmospheric modelling. In Section~\ref{section:photometry} we discuss the analysis of the existing TESS data. Finally, we discuss the presence and lack of tidal signals and present a toy model to explain our observations in Section~\ref{section:mode_tracking} before concluding in Section~\ref{section:conclusions}.

\section{UVES spectroscopy of U~Gru}
\label{section:spectra}

To characterise the components and the orbit of the system, we obtained high-resolution spectroscopy with the Ultraviolet and Visual {\'E}chelle Spectrograph \citep[{\sc uves} on UT2@VLT; ][]{Dekker2000} via a Director's Discretionary Time proposal (ID: 0104.D-0209; PI: C. Johnston). Our observations utilized the standard DIC-2 blue arm setting, which has both a blue ($\lambda\sim300-500$~nm; $\lambda_{c}=437$~nm; R$\sim$80\,000) and red ($\lambda\sim 420 - 1100$~nm; $\lambda_{c}=760$~nm; R$\sim$110\,000) detector equipped on {\sc uves}. All observations were taken using a 1\arcsec slit with a 600 sec exposure time. We obtained 12 epochs over a three month period, which are distributed to maximize coverage across the orbital period of the U Gru binary system. The observations are briefly summarized in Table~\ref{tab:spec_table}.

The spectra were extracted, corrected, and calibrated with the standard version of the {\sc eso uves} pipeline \citep{Sacco2014} and can be obtained directly from the ESO  archive\footnote{\url{http://archive.eso.org/wdb/wdb/adp/phase3_spectral/form?collection_name=UVES}}. Further normalisation of the reduced spectra to the local continuum was performed by fitting a 3rd order polynomial through a set of manually selected continuum points. Barycentric velocity and time corrections were applied to the spectra.

\begin{table}[ht]
    \centering
    \caption{Barycentric Julian Date, orbital phase, and S/N at 4500\AA  of the UVES observations.}
    \begin{tabular}{lcc}
        \hline
        BJD & Phase & S/N      \\
        d   &       &  \\
        \hline
        
        2458707.80299  &  0.06  & 70  \\
        2458679.74695  &  0.14  & 114 \\
        2458698.73534  &  0.24  & 88  \\
        2458700.82038  &  0.35  & 123 \\
        2458672.73848  &  0.42  & 103 \\
        2458674.75670  &  0.49  & 103 \\
        2458661.73705  &  0.57  & 98  \\
        2458678.81785  &  0.65  & 130 \\
        2458633.88509  &  0.76  & 112 \\
        2458699.80944  &  0.81  & 112 \\
        2458720.68674  &  0.91  & 115 \\
        2458675.68874  &  0.99  & 37  \\

        \hline
    \end{tabular}
    \label{tab:spec_table}
\end{table}

\subsection{Least-Squares Deconvolution}\label{sec:LSD}

The first step of our spectroscopic analysis involves computing the least-squares deconvolution \citep[LSD;][]{Donati1997} profiles with a customised version of the LSD algorithm \citep{Tkachenko2013}. An LSD profile is an average line profile that is characterised by a significantly enhanced signal-to-noise ratio (S/N) compared to individual absorption lines in the stellar spectrum. Additionally, the LSD profile captures any type of variability that is common to all spectral lines involved in the calculation. In particular, the LSD method provides an efficient way of detecting line profile variations due to binarity and/or intrinsic variability of either of the binary components in the form of rotational modulation and/or stellar pulsations \citep[e.g., ][]{Tkachenko2013,Johnston2021}.

\begin{figure}
    \centering
    \includegraphics[width=0.95\columnwidth]{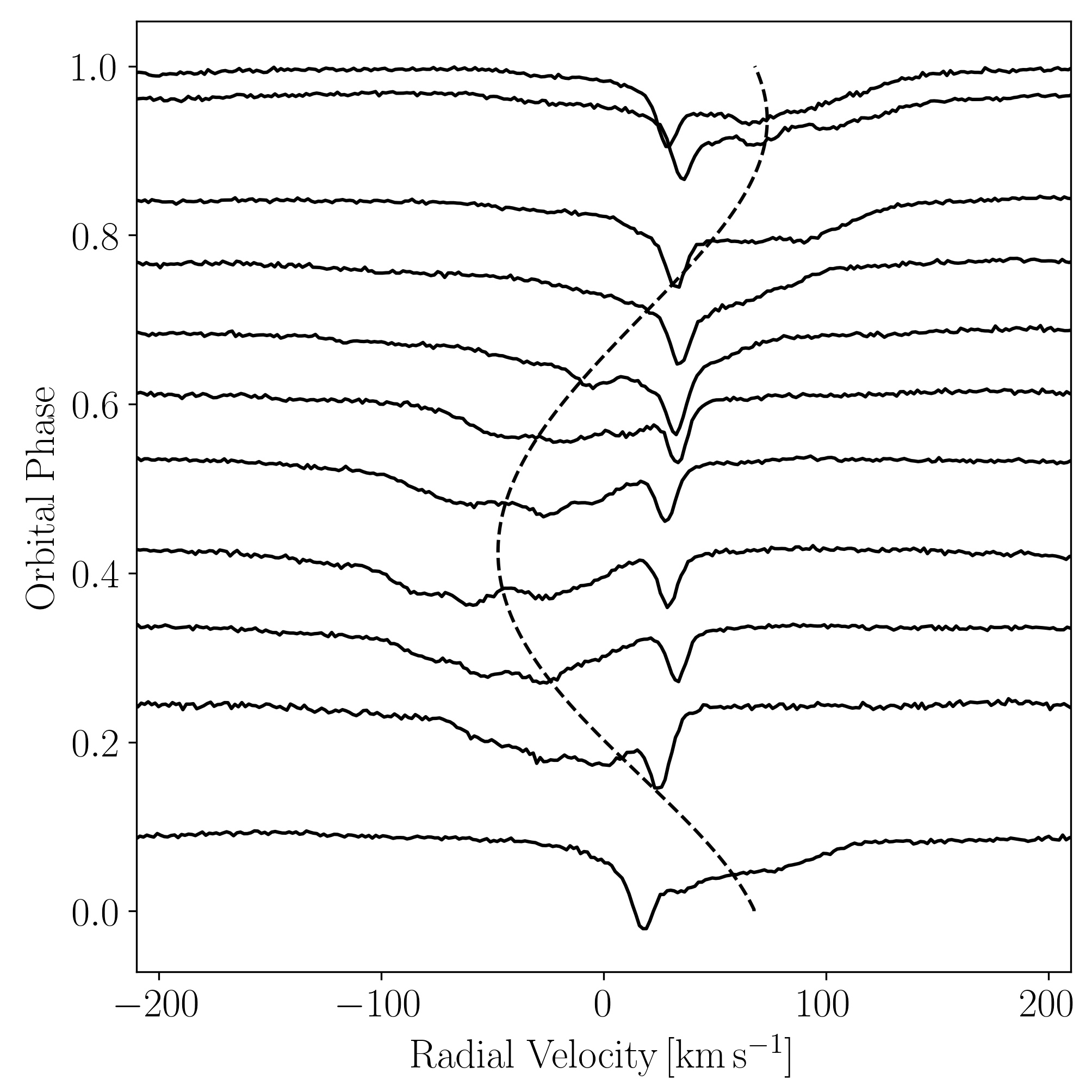}
    \caption{The least-squares deconvolution profiles of U~Gru computed from 11 blue arm {\sc uves} spectra. The orbital phase resolved spectral contribution of the broad-lined primary component is indicated with the dashed line.}
    \label{fig:phased_lsd}
\end{figure}

To compute the LSD profiles, we used multiple synthetic line lists for an A-type star ($T_{\rm eff}=8\,000,~\log g=4.0~{\rm dex},~{\rm [M/H]=0}$), a G-type star ($T_{\rm eff}=6\,000,~\log g=4.0~{\rm dex},~{\rm [M/H]=0}$), as well as for a K-type star ($T_{\rm eff}=5\,000,~\log g=4.0~{\rm dex},~{\rm [M/H]=0}$). The line lists were constructed from the VALD database \citep{Kupka2000}, covering a wavelength range of 4\,000-7\,000~\AA, excluding Balmer lines. Although we compute LSD profiles from the spectra obtained with both the blue and red arms of the {\sc uves} instrument, below we discuss the results obtained for the blue arm only. We do this to avoid repeating the results given the strong similarities obtained from the red part of the spectrum. Additionally, we do not use the spectrum taken during primary eclipse as the S/N is low and we do not see the signature of the primary. The orbital phase folded LSD profiles for 11 spectra from the {\sc UVES} blue arm are shown in Fig.~\ref{fig:phased_lsd}. We note that the systemic velocity has not been removed from the observations before we calculate the LSD profiles. Two components are clearly visible in the LSD profiles of the blue arm. One component is broadened and shows intrinsic line-profile variability throughout the motion along the orbit. The other component reveals a narrow-lined profile which appears stationary over the orbital period. We identify the broad-lined profile as the rejuvenated A-type primary (i.e. the $\delta$~Scuti pulsator), and identify the narrow-lined profile as a potential slowly rotating tertiary component in the system due to its negligible radial velocity (RV) variability. The cool faint secondary component is not firmly detected in either of the two arms of the obtained {\sc uves} spectra.

\subsection{Spectral disentangling}

Spectroscopic observations of multiple star systems are comprised of the composite light of all the $N$ components in the system. The analysis of such a composite spectrum can be done by directly fitting $N$ components to the spectrum with the appropriate RV shifts and light factors, or by separating the components using techniques such as spectral disentangling. The method of spectral disentangling allows one to simultaneously optimise for orbital elements of the system and separate the individual spectra of the stars forming the system \citep{Simon1994,Hadrava1995}. For the purpose of spectral disentangling, we employ the {\sc FDBinary} software package \citep{Ilijic2004} using 11 out of the total 12 acquired blue arm spectra of U~Gru. In particular, we remove the spectrum taken at the time of the eclipse of the more luminous primary component due to the strong variations in the spectral line depths caused by the near total blockage of light from the primary component. We perform the spectral disentangling in the wavelength region between the $H_{\gamma}$ and $H_{\beta}$ Balmer profiles spanning 4\,400-4\,800~\AA.

In the first step, we perform the spectral disentangling in a triple system mode, primarily in an attempt to detect signatures of the faint secondary component. However, similar to the results obtained with the LSD profiles, no trace of the cool secondary component can be found with the spectral disentangling technique. We therefore proceed with the spectral disentangling in a `fake binary mode' where the narrow-lined tertiary component is constrained to have long periods such that it shows little to no RV variability over the 3-month time span of our {\sc uves} observations. The broad-lined primary component is assumed to be on a circular orbit and is otherwise unconstrained. This is supported from the initial TESS photometry and preliminary binary modelling by \citet{Bowman2019}. Both of these assumptions are largely consistent with the scenario arising from the properties of the LSD profiles outlined in Section~\ref{sec:LSD}. From this, we find the RV semi-amplitude of the more luminous primary component to be $K_{\rm 1} = 61.5\pm 0.7$~km\,s$^{-1}$ and find no significant RV variability associated with the tertiary component. 

\begin{figure}
\centering
\includegraphics[width=0.95\linewidth]{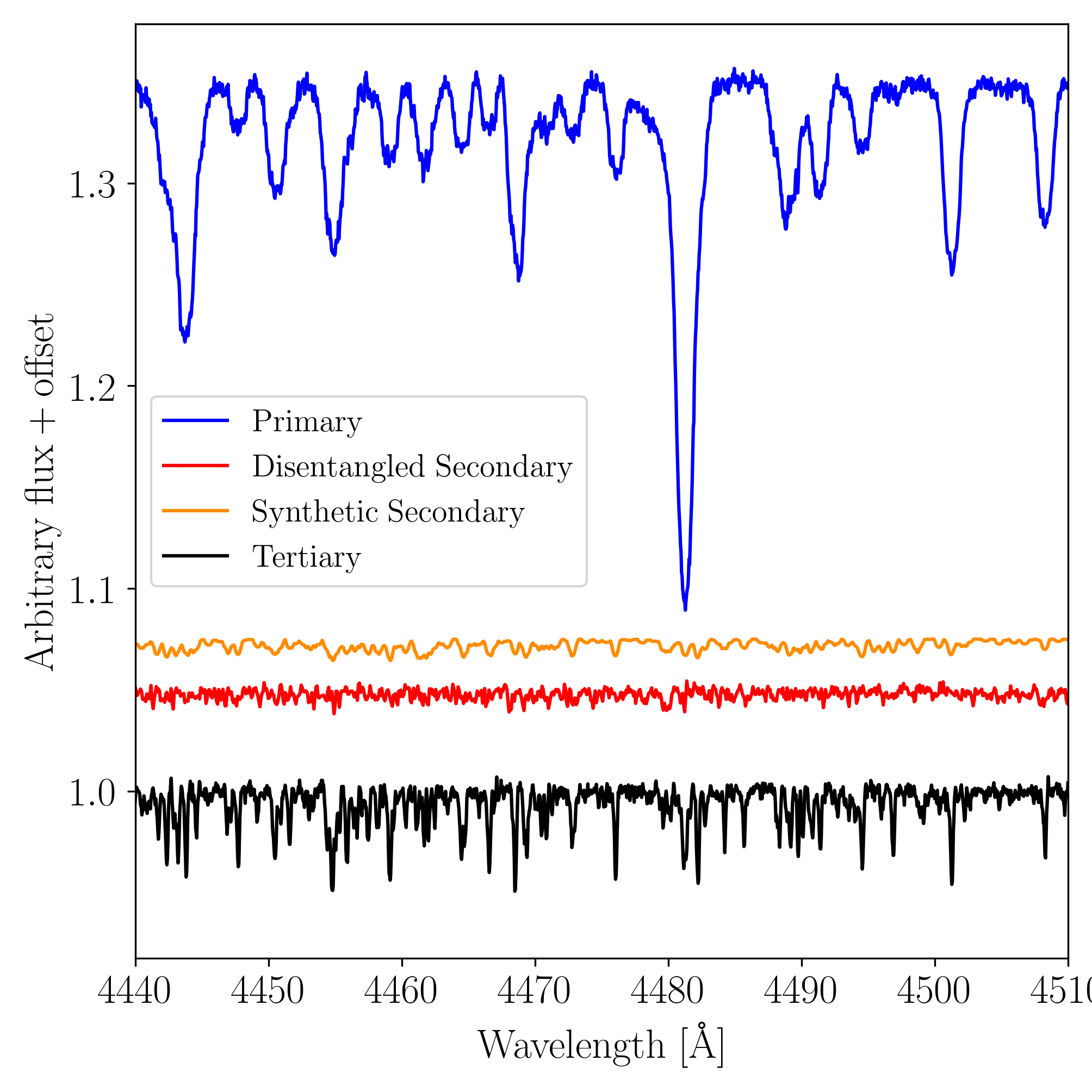}
    \caption{Disentangled spectra of the primary (blue) and tertiary (black) components of the U~Gru system. A representative synthetic spectrum corresponding to $T_{\rm eff}=5500$~K and $\log\,g=3.5$~dex and diluted to the level of 2\% (in continuum units) is shown for comparison to the secondary component (red). Note that a constant vertical shift have been applied to all the spectra for visibility purposes. 
    }
\label{fig:SPD_tertiary_secondary}
\end{figure}

Using the orbital solution and the disentangled spectra of the primary and tertiary components from the `fake binary mode', we shift these disentangled spectra by their individual RVs corresponding to the times of each of the {\sc uves} observations and then subtract each shifted disentangled spectrum from the originally observed composite spectra. This way we obtain a time series of eleven residual spectra in which spectral contributions of the primary and tertiary components have been eliminated. In the next step, we explore these residual spectra with the {\sc FDBinary} algorithm in a spectroscopic single-lined binary (SB1) mode in an attempt to search for extremely weak signatures of the cool secondary component. This exercise reveals a tentative spectrum of a cool star whose RV semi-amplitude $K_{\rm 2}$ is found to be between 195 and 200~km\,s$^{-1}$. Figure~\ref{fig:SPD_tertiary_secondary} shows the disentangled spectra of the primary (blue) and tertiary (black) components of the U~Gru system, as well as the tentative disentangled spectrum of the secondary (red) alongside a synthetic spectrum computed for $T_{\rm eff}=5500$~K and $\log\,g=3.5$~dex (orange) and arbitrarily diluted to the level of 2\% (in continuum units) for comparison. We note that the disentangled secondary spectrum has been binned by a factor of two to increase the signal to noise. Although we do not claim a firm detection of the secondary component, we do identify some similarities between the disentangled secondary spectrum and the synthetic spectrum in Fig.~\ref{fig:SPD_tertiary_secondary}.

\begin{figure}
\centering
\includegraphics[width=0.95\linewidth]{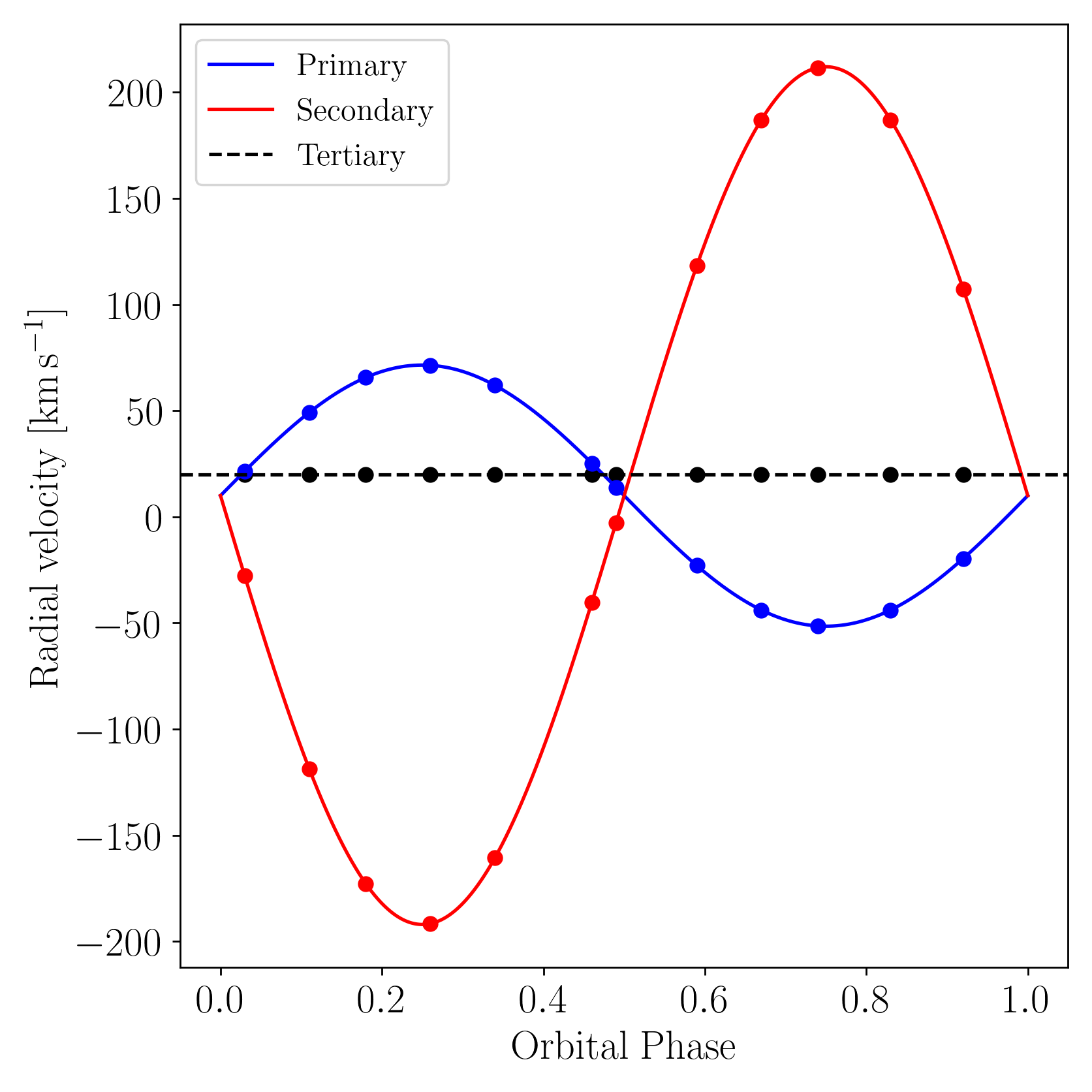}
    \caption{Orbital solution for the primary (blue), secondary (red), and tertiary (black) components of U~Gru obtained with the method of spectral disentangling. Note that the symbols do not represent actual RV measurements. Instead, they indicate the distribution of the observed spectra over the orbital cycle of the inner binary system, while the orbital RV curve itself is computed from the best fit orbital elements.}
\label{fig:SPD_orbit}
\end{figure}

As the last exercise, we fix $K_{\rm 1}$ and set the initial guess for $K_{\rm 2}$ according to our previous findings, and once again perform spectral disentangling in the SB3 triple system mode. The inner binary orbit is assumed to be circular and we allow for small RV variability for the narrow-lined tertiary component. We find the RV semi-amplitude of the secondary component to be $K_{\rm 2}$=202$\pm$3~km\,s$^{-1}$ while the disentangled spectra of all three components do not show any significant differences with respect to the solutions described above. We find that the tertiary RV semi-amplitude tends towards 0~km\,s$^{-1}$ in this solution. Together with $K_{\rm 1}$ the above value of $K_{\rm 2}$ suggests a mass ratio of $q=0.304\pm0.005$, and $\sim$2.7~$M_{\odot}$ and $\sim$0.8~$M_{\odot}$ for the mass of the primary and secondary components, respectively, assuming an inclination of $i=90^{\circ}$. All these values are consistent with the system configuration typical for Algols and the mass of the primary is consistent with its late-A spectral type. We note that given the lack of a robust detection of the secondary, we do not attempt a full binary modelling in this work as any solution would be degenerate without an independent constraint on the mass-ratio and semi-major axis of the system. 

Figure~\ref{fig:SPD_orbit} shows the orbital RV curves for the primary and secondary components of U~Gru computed from the corresponding best fit orbital elements. Because the method of spectral disentangling by-passes the step of RV determination and instead solves for the orbital elements directly, the symbols shown in Fig.~\ref{fig:SPD_orbit} do not represent the actual RV measurements of the primary, secondary, and tertiary components, respectively. Instead, the symbols represent the RVs computed from the best fit orbital solution at the times of our {\sc uves} observations. We show this to demonstrate the orbital phase coverage achieved with our spectroscopic observations. 

\subsection{Atmospheric parameters}

We employ the Grid Search in Stellar Parameters \citep[{\sc gssp};][]{tkachenko2015} software package to infer atmospheric parameter values of the primary and tertiary components from their disentangled spectra. We do not perform spectrum analysis of the faint secondary component owing to the low S/N in its disentangled spectrum. The {\sc gssp} package is employed in its `unconstrained fitting' mode that allows us to analyse the disentangled spectra of stars as if they were single. Even though {\sc gssp} offers an option to fit the disentangled spectra of binary (or triple system) components simultaneously and taking into account wavelength dependence of the components' light contributions, we do not opt for that particular mode in our analysis. The main reason is because we expect the secondary component to potentially still contaminate the spectrum of the primary component. In turn, the uncertainty in the disentangled spectrum of the primary component propagates into the results obtained for the tertiary component when the two spectra are coupled methodologically. 

In our spectrum analysis, we explore different combinations of free and fixed atmospheric parameters, of which we consider in total six for each of the components: effective temperature $T_{\rm eff}$, surface gravity $\log\,g$, microturbulent velocity $\xi$, projected rotational velocity $v\,\sin\,i$, metallicity [M/H] (assuming an \citet{Asplund2005} solar composition), and light dilution factor $\ell f_{\rm i}$. Immediately, we find that although the light dilution factor is well constrained for the tertiary component, no (realistic) optimal value can be obtained for the primary component. Therefore, we start with the atmospheric analysis of the tertiary component and then proceed with the analysis of the primary component where its light dilution factor $\ell f_{\rm 1}$ is fixed according to the relation $\ell f_{\rm 1}$ = 1 - $\ell f_{\rm 3}$. This relation assumes that the secondary component is not present in the system. Instead, in the final stage of our analysis, we re-determine atmospheric parameters of the primary and tertiary components allowing for a few percent variations of the light factors $\ell f_{\rm 1}$ and $\ell f_{\rm 3}$ from their optimal values. We do not consider variations larger than a few percent because that is what we expect for the light contribution from the faint secondary component. We account for deviations of the atmospheric parameters of the primary and tertiary components from their optimal values due to variable $\ell f_{\rm 1}$ and $\ell f_{\rm 3}$ in the final reported parameter uncertainties.

Our spectrum analysis show that the most stable spectroscopic solution is achieved when the $\xi$ and [M/H] parameters are fixed to 2~km~s$^{-1}$ and 0.0~dex (solar value), respectively. We note that the metallicity [M/H] in the atmospheric spectroscopic analysis refers to a joint contribution of all elements heavier than helium. Otherwise, we find a statistically significant difference in the metallicity of the two stars. This scenario is not expected in a gravitationally bound binary (or triple) system where all components are formed from the same natal cloud, despite the past history of binary interaction. However, similar to the case with the light factors, we do account for the effects of metallicity and microturbulent velocity in the finally reported uncertainties for $T_{\rm eff}$, $\log\,g$, and $v\,\sin\,i$. The final atmospheric parameters (and their uncertainties) of the primary \citep[A1V;][]{Pecaut2013} and tertiary (F3V) component of U~Gru are reported in Table~\ref{tab:gssp_params}. Given the surface gravity and mass range of the primary as derived above, the primary is rotating at less than 20\% its critical rate.

 \begin{table}[t]
     \centering
     \caption{Atmospheric parameters of the primary and tertiary components of U~Gru. See text for details.}
     \begin{tabular}{lccc}
         \hline
         \multirow{2}{*}{Parameter} & \multirow{2}{*}{Unit} & \multicolumn{2}{c}{Component} \\
          & & Primary & Tertiary \\
         \hline
         $T_{\rm eff}$ & K   &  9000$\pm$550 & 6710$\pm$450 \\
         $\log\,g$ & dex   &  4.40$\pm$0.35 & 4.35$\pm$0.5 \\
         $\xi$ & km~s$^{-1}$   &  \multicolumn{2}{c}{2.0 (fixed)} \\
         $v\,\sin\,i$ & km~s$^{-1}$   &  59$\pm$5 & 6.8$\pm$1.9 \\         
         ${\rm [M/H]}$ & dex   &  \multicolumn{2}{c}{0.0 (fixed)} \\
         $f_{\rm i}$ & dex  & 0.94 (fixed) & 0.06$\pm$ 0.02\\
         \hline
     \end{tabular}
     \label{tab:gssp_params}
 \end{table}

\section{TESS photometry}
\label{section:photometry}
\begin{figure}
\centering
\includegraphics[width=0.95\linewidth]{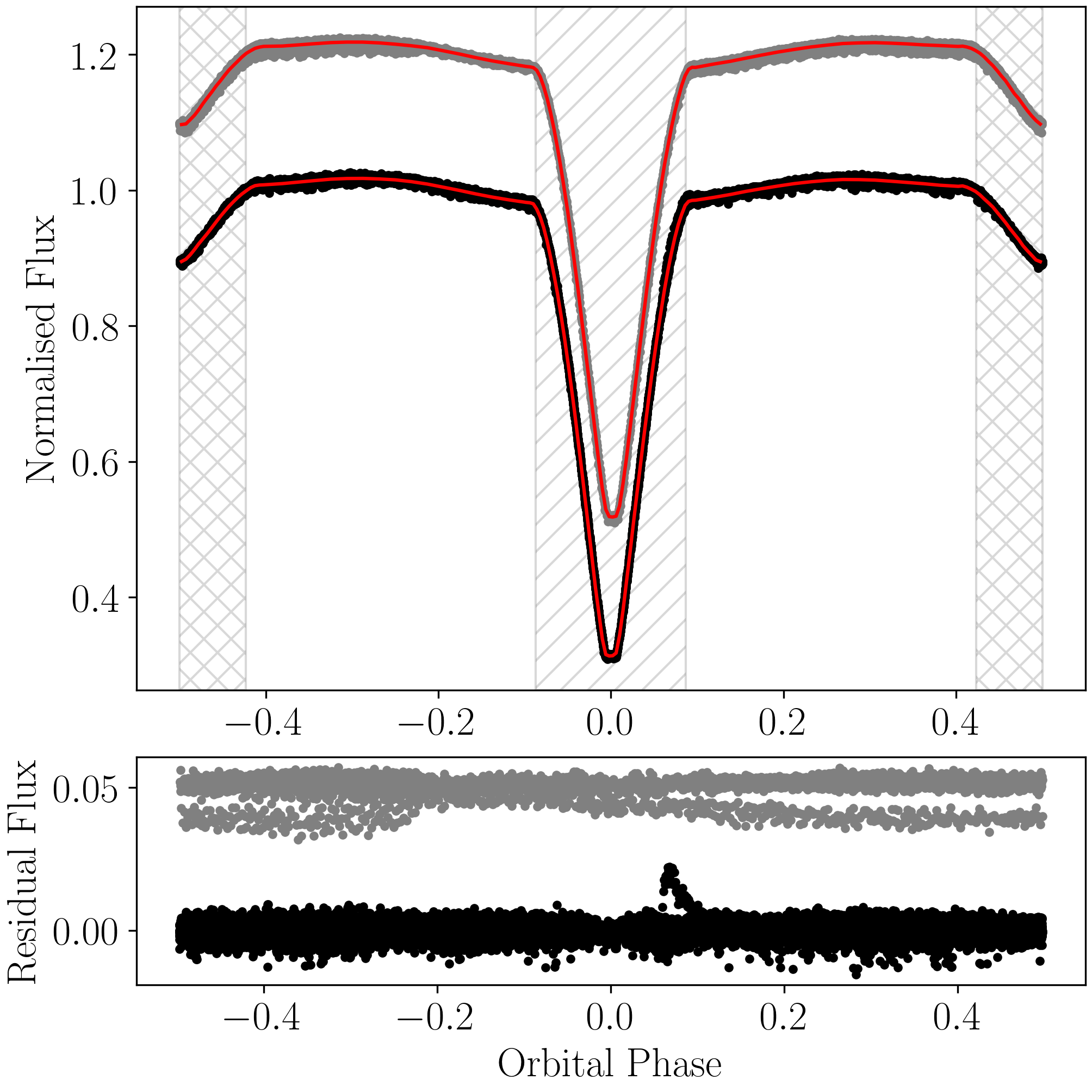}
    \caption{({\it Top}):  Phase folded light curve of U~Gru for sectors 1 (black) and 28 (grey) with the interpolated binary model in red. Grey hatching denotes phases of primary (forward hatching) and secondary (cross hatching) eclipse. ({\it Bottom}): Residuals after subtraction of interpolation binary model for sectors 1 (black) and 28 (grey).}
\label{fig:phased_lightcurves}
\end{figure}

\begin{figure}
\centering
\includegraphics[width=0.95\linewidth]{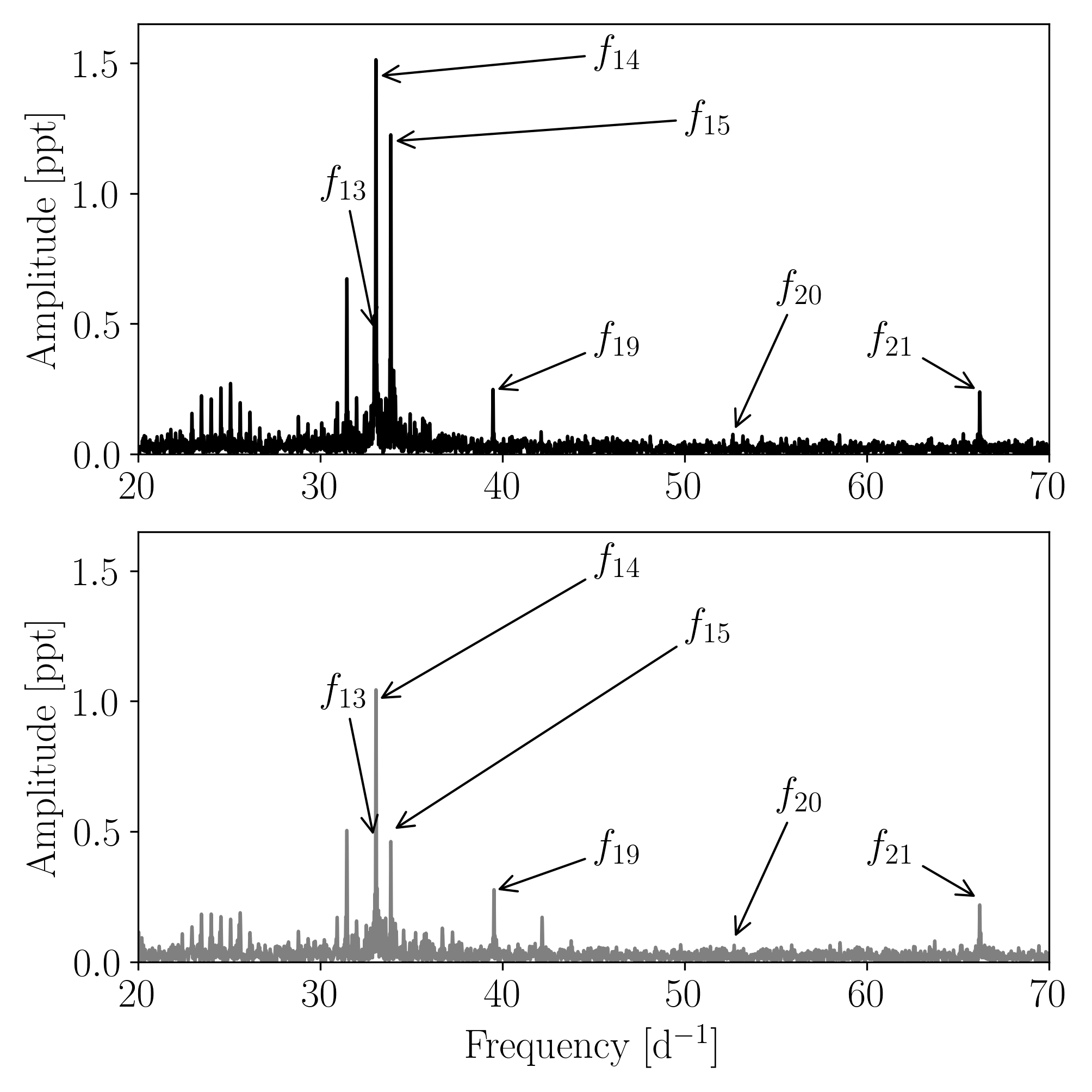}
    \caption{({\it Top}): Lomb-Scargle periodogram of binary subtracted residuals for sector 1. 
             ({\it Bottom}): Lomb-Scargle periodogram of binary subtracted residuals for sector 28.
             Labelled frequencies denote frequencies that are tracked in 
             Section~\ref{section:tidal_effects} and shown in orange in Fig.~\ref{fig:echelle}}
\label{fig:compare_periodograms}
\end{figure}

U~Gru was observed by TESS in two sectors, separated by roughly two years. During the first sector, U~Gru was observed for 27.9~d in 2-min cadence mode, whereas it was observed for 25.2~d at a 10-min cadence in the full-frame image (FFI) data during sector 28. We find no indication of the tertiary object in either sector of the TESS photometry. Using the 2-min cadence sector 1 data, \citet{Bowman2019} reported 22 significant pulsation frequencies after subtracting a multi-harmonic model to remove the eclipsing binary contribution. However in this work we remove the strictly periodic eclipsing signal by using an interpolated model constructed from the orbital phase-folded light curve. In brief, we phase fold the original light curve on the orbital period, and then bin the phased light curve. We then construct an interpolated model from the phase-binned data and subtract that model from the original light curve. The phase folded light curves from sectors 1 and 28 are shown in black and grey in the top panel of Fig.~\ref{fig:phased_lightcurves}. The interpolated models are shown in red, and the residuals are shown in the bottom panel. The Lomb-Scargle (LS) periodograms \citep{Lomb1976,Scargle1982,Press1989,Astropy2018} of the interpolation model subtracted residuals are shown in the black (sector 1) and grey (sector 28) in Fig.~\ref{fig:compare_periodograms}. As can be seen, the series of frequencies separated by (multiples of) the orbital frequency  are still seen in the residuals from sector 28, suggesting that the underlying phenomenon is stable across the two years separating the data sets. Notably, the amplitudes observed in the periodogram of the sector 28 residuals are generally lower than those observed in the sector 1 residuals. This is likely caused by the change in photometric sampling rate (two min to 10 min) between the sectors. There does appear to be an additional amplitude change for the second highest amplitude frequency. Intrinsic pulsation mode amplitude variability over multiple years is commonly observed in $\delta$~Scuti pulsators and is attributed to mode coupling or changes to the internal stellar structure \citep{Bowman2016}.

\begin{figure}
\centering
\includegraphics[width=0.95\linewidth]{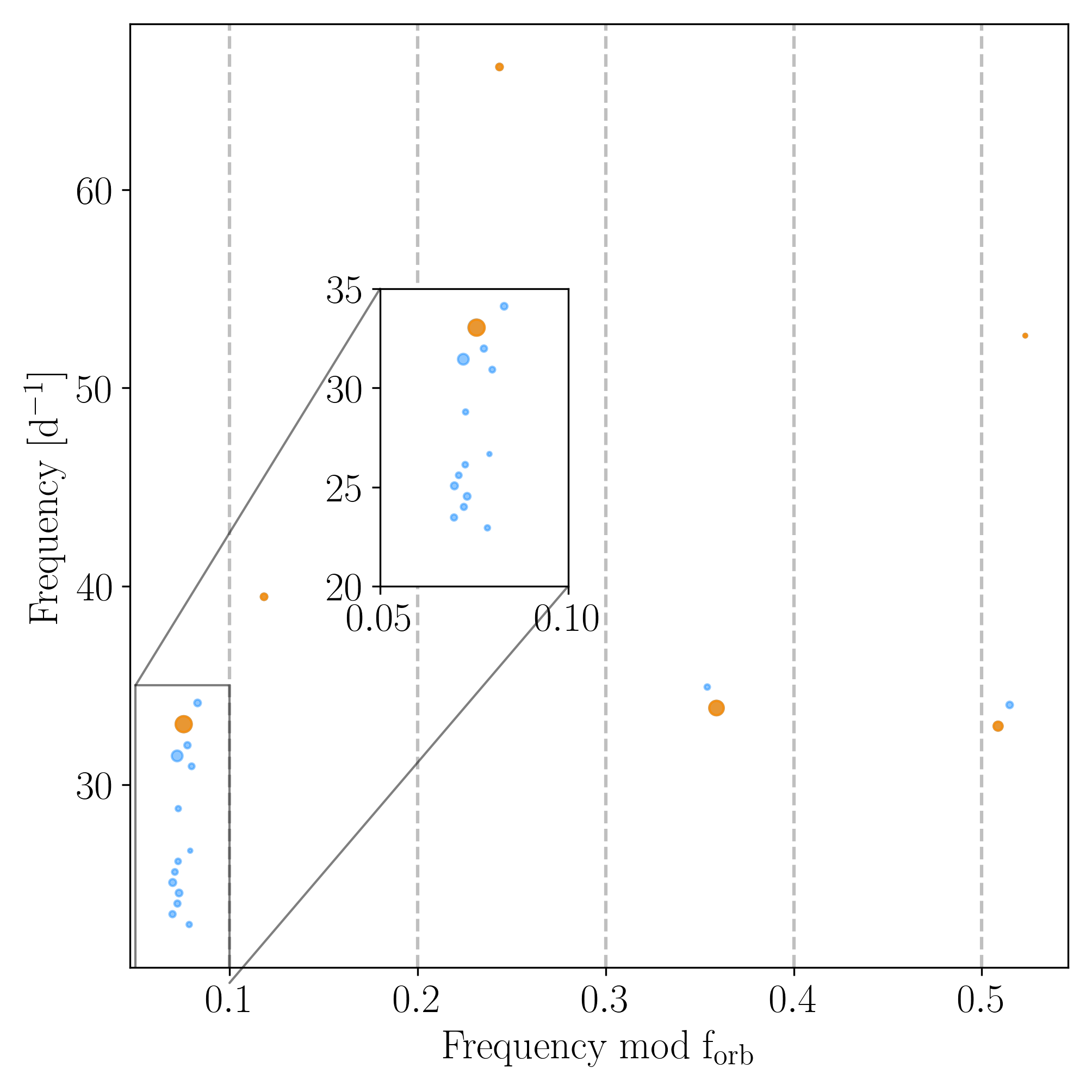}
    \caption{{\'E}chelle diagram of the significant frequencies extracted from the residuals of sector 1 folded over the orbital period. The marker size scales with amplitude. Orange markers denote frequencies labelled in Fig.~\ref{fig:compare_periodograms} and tracked in Section~\ref{section:tidal_effects}}
\label{fig:echelle}
\end{figure}

We conduct an iterative pre-whitening procedure on the residuals of the sector 1 data using the {\sc pythia} code\footnote{\url{https://github.com/colej/pythia}} in order to remove the remaining periodic signals. We identify the signal to be removed at each iteration as the peak with the highest S/N calculated via a 1~d$^{-1}$ window surrounding the frequency. The signal in question is modelled with a sinusoid, where the optimised parameters are taken as the {\it maximum a posteriori} (MAP) non-linear least-squares fit to the light curve. We perform the optimisation using the {\sc pymc3} \citep{salvatier2016} and {\sc exoplanet} \citep{ForemanMackey2021} {\sc python} modules. We iterate until we extract all signals with S/N > 4, and then re-calculate the final model with all frequencies as free parameters. We calculate the final S/N with respect to the residual light curve. We find 21 significant frequencies, which are reported in Table~\ref{tab:ipw_results}. The phases are calculated with respect to $t_0=2458325.93900$, which is the reference epoch of primary eclipse first reported by \citealt{Bowman2019}. We report uncertainties on the frequencies, amplitudes, and phases according to \citet{Montgomery1999} accounting for the correlated nature of the data according to \citet{Schwarzenberg2003}. We recover 8 fewer frequencies that are split by the orbital frequency than those originally reported in \citet{Bowman2019}. Five of these modes had a S/N < 4 reported by \citet{Bowman2019}, whereas all eight were found to be formally insignificant in our analysis. We find seven additional significant frequencies that \citet{Bowman2019} did not recover in their analysis. The differences in extracted frequencies is likely caused by the different methods used to subtract the binary signal and the different methods for calculating the noise level around a given frequency. We construct and show an {\'E}chelle diagram of our extracted frequencies folded over the orbital frequency in Fig.~\ref{fig:echelle}, which shows a clear series of frequencies split by (multiples of) the orbital frequency.

\begin{table*}
    \centering
    \caption{Optimised frequencies, amplitudes, and phases of the significant frequencies identified via iterative pre-whitening. We also list the S/N of each extracted frequency, as well as its value modulo the orbital frequency. The zero-point in time was set to $t_0=2458325.93900$.}
    \label{tab:ipw_results}
    \begin{tabular}{llcccc}
    \hline
                 & Frequency           & F\,mod\,$f_{\rm orb}$  & Amplitude     & Phase            & S/N  \\
                 & d$^{-1}$            &                        &  ppt          & rad              &      \\
    \hline
        $f_{01}$ & $22.945  \pm0.006$  & $0.079  \pm0.006$  & $0.16\pm0.05$ & $1.06  \pm0.02$  & 5.5  \\
        $f_{02}$ & $23.468  \pm0.004$  & $0.070  \pm0.004$  & $0.22\pm0.05$ & $1.72  \pm0.02$  & 8.2  \\
        $f_{03}$ & $24.002  \pm0.004$  & $0.072  \pm0.004$  & $0.21\pm0.05$ & $1.67  \pm0.02$  & 9.7  \\
        $f_{04}$ & $24.535  \pm0.004$  & $0.073  \pm0.004$  & $0.26\pm0.05$ & $1.53  \pm0.01$  & 10.7 \\
        $f_{05}$ & $25.063  \pm0.003$  & $0.070  \pm0.003$  & $0.28\pm0.05$ & $1.76  \pm0.01$  & 10.2 \\
        $f_{06}$ & $25.596  \pm0.005$  & $0.071  \pm0.005$  & $0.19\pm0.05$ & $1.50  \pm0.02$  & 7.2  \\
        $f_{07}$ & $26.130  \pm0.005$  & $0.073  \pm0.005$  & $0.17\pm0.05$ & $1.67  \pm0.02$  & 7.5  \\
        $f_{08}$ & $26.670  \pm0.009$  & $0.080  \pm0.009$  & $0.10\pm0.05$ & $1.23  \pm0.04$  & 4.6  \\
        $f_{09}$ & $28.789  \pm0.006$  & $0.073  \pm0.006$  & $0.15\pm0.05$ & $-1.14 \pm0.02$  & 5.2  \\
        $f_{10}$ & $30.923  \pm0.005$  & $0.080  \pm0.005$  & $0.19\pm0.05$ & $1.69  \pm0.02$  & 5.6  \\
        $f_{11}$ & $31.447  \pm0.001$  & $0.072  \pm0.001$  & $0.67\pm0.05$ & $-1.70 \pm0.01$  & 17.4 \\
        $f_{12}$ & $31.984  \pm0.004$  & $0.078  \pm0.004$  & $0.22\pm0.05$ & $1.97  \pm0.02$  & 5.9  \\
        $f_{13}$ & $32.947  \pm0.002$  & $0.509  \pm0.002$  & $0.47\pm0.05$ & $1.274 \pm0.01$  & 6.6  \\
        $f_{14}$ & $33.046  \pm0.001$  & $0.076  \pm0.001$  & $1.49\pm0.05$ & $-1.51 \pm0.01$  & 21.8 \\
        $f_{15}$ & $33.861  \pm0.001$  & $0.359  \pm0.001$  & $1.20\pm0.05$ & $-1.30 \pm0.01$  & 26.9 \\
        $f_{16}$ & $34.017  \pm0.004$  & $0.515  \pm0.004$  & $0.24\pm0.05$ & $-2.28 \pm0.02$  & 5.2  \\
        $f_{17}$ & $34.117  \pm0.004$  & $0.083  \pm0.004$  & $0.25\pm0.05$ & $1.46  \pm0.02$  & 6.5  \\
        $f_{18}$ & $34.919  \pm0.006$  & $0.354  \pm0.006$  & $0.16\pm0.05$ & $2.37  \pm0.02$  & 4.1  \\
        $f_{19}$ & $39.470  \pm0.004$  & $0.118  \pm0.004$  & $0.24\pm0.05$ & $0.73  \pm0.02$  & 11.1 \\
        $f_{20}$ & $52.640  \pm0.009$  & $0.520  \pm0.009$  & $0.08\pm0.05$ & $1.29  \pm0.05$  & 4.0  \\
        $f_{21}$ & $66.184  \pm0.004$  & $0.244  \pm0.004$  & $0.24\pm0.05$ & $1.32  \pm0.02$  & 10.8 \\
    \hline
        \end{tabular}

\end{table*}

\section{Amplitude and phase tracking}
\label{section:mode_tracking}
\subsection{Tidal effects}
\label{section:tidal_effects}

In all of the confirmed tidally tilted pulsating stars known to date, the tidal splitting has resulted in a (generally) symmetric multiplet, split by (multiples of) the orbital frequency. In the case of U~Gru, however, \citealt{Bowman2019} identified a series of frequencies that did not appear to be centred on a particular oscillation mode, posing a challenge for interpreting this phenomenon as the result of a tidally tilted pulsation axis. 

\begin{figure*}
\centering
\includegraphics[width=0.95\linewidth]{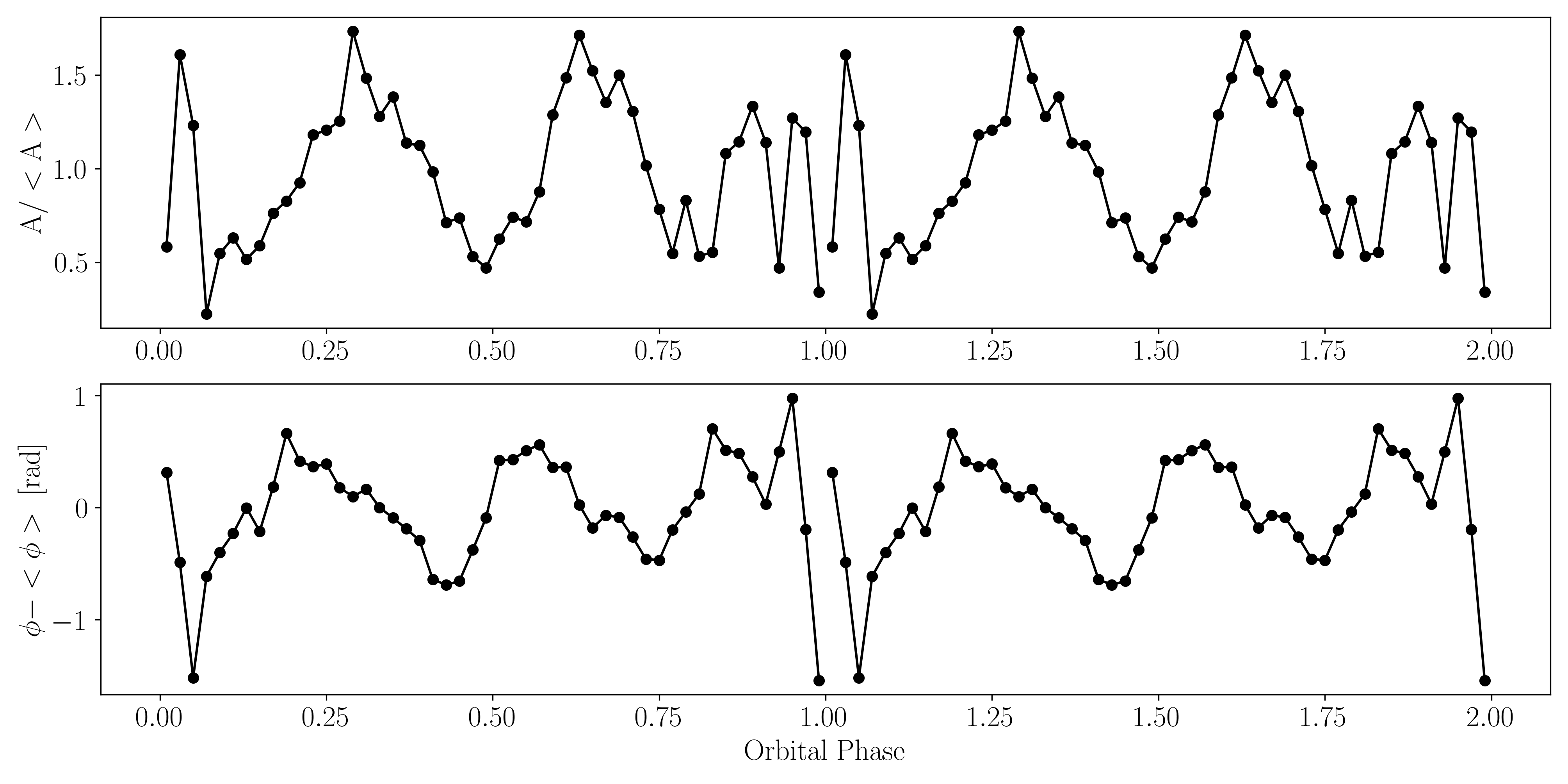}
    \caption{({\it Top}): Tracked mode amplitude for $f_{14}=33.046$~d$^{-1}$ as a function of orbital phase. ({\it Bottom}): Tracked relative mode phase for $f_{14}=33.046$~d$^{-1}$ as a function of orbital phase. Phase values of 0, 1 and 2 correspond to primary eclipse, and phase values of 0.5 and 1.5 correspond to secondary eclipse.  }
\label{fig:obs_modulations}
\end{figure*}

In this work we re-visit the pulsation analysis of U~Gru, and track the amplitude and phases of the several frequencies extracted from iterative pre-whitening. We divide the {\'E}chelle diagram into distinct sections (e.g., 0-0.1~d$^{-1}$, 0.1-0.2~d$^{-1}$, etc.) as shown by the grey dashed lines in Fig.~\ref{fig:echelle}, and track the highest amplitude mode for each section of the {\'E}chelle diagram, e.g. $f_{13}$, $f_{14}$, $f_{15}$, $f_{19}$, and $f_{21}$. We also consider$f_{20}$ to be independent and track it. These modes are labelled in Fig.~\ref{fig:compare_periodograms} and denoted by orange markers in Fig.~\ref{fig:echelle}. To do this, we remove all other significant frequencies  from the other sections of the {\'E}chelle diagram before dividing the orbital phase-folded light curve into 50 equally sized bins. We fit a sinusoid with a fixed frequency to each bin and record the amplitude and phase following \citet{VanReeth2022}. The tracked amplitudes and phases are shown in Fig.~\ref{fig:obs_modulations}, with orbital phase values of 0, 1 and 2 corresponding to primary eclipse, and orbital phase values of 0.5 and 1.5 corresponding to secondary eclipse. In order to compare the modulations for each frequency, we consider the relative amplitude and phase modulations. As such, we divide the tracked amplitude at each phase bin by the average amplitude $<A>$ across all orbital phase-bins considered. Similarly, we subtract the average mode phase $<\phi>$ for each frequency. Only the highest amplitude pulsation at $f_{14}$=33.046~d$^{-1}$ shows amplitude and phase modulation across the orbit, as shown in Fig.~\ref{fig:obs_modulations}. Additionally, we note that the remaining frequencies that we tracked amplitudes and phases for show amplitude minima during primary eclipse.

We find clear evidence that the amplitude and phase of the p~mode at $f_{14}=33.046$~d$^{-1}$ is modulated over the binary orbit. We note that the amplitude modulation displays an apparent minima during secondary eclipse, with maxima occurring at $\frac{1}{3}$ phases of the orbit. The three observed amplitude maxima are difficult to explain with the currently proposed tidal tilting framework as a tilted mode with degree $\ell$ should produce 2$\ell$ amplitude maxima across the binary orbit \citep{Fuller2020}. Furthermore, we observe phase modulations with a $\pi / 2$~rad variation, similar to those observed in distorted modes in roAp stars \citep[e.g., ][]{Holdsworth2016} and the g-mode pulsating binary V456~Cyg \citep{VanReeth2022}. Finally, we observe a complex behaviour of the amplitude and phase modulation at phases surrounding primary eclipse. Following the oblique pulsator model assumed for the case of tidal trapping, we should observe a symmetric multiplet split by the orbital frequency from the intrinsic pulsation frequency as well as phase variations with values of $0$, $\pi$, or $2\pi$. This model cannot explain either the full series of frequencies split by (multiples of) the orbital frequency nor the observed phased modulations of the dominant p mode in U~Gru.

\subsection{Eclipse mapping}
\label{section:eclipse_mapping}
Photometric observations of distant, point source stars are the total integrated surface brightness of the projected stellar surface at a given time. Stellar pulsations induce brightness variations in time by modifying the surface brightness distribution according to the geometry of the pulsation modes. In the case of an eclipsing binary, the projected stellar disk is partially (or totally) obscured, resulting in a time-varying amount of the stellar disk that is observed. This produces a time-dependent modification to the partial cancellation of a varying mode over the stellar surface \citep{Aerts2010}. U~Gru exhibits total eclipses of the primary, wherein the secondary component obscures nearly the entire visible disk of the pulsating A-type primary. Therefore, there is an additional modulation of the observed pulsation amplitude as the disk of the pulsating star is partially obscured during ingress to primary eclipse, blocked totally during conjunction, and then partially obscured again during egress from primary eclipse. This process creates a complicated modulation of the observed pulsation amplitude in time as part of the variable surface is preferentially obscured. This phenomenon is sometimes referred to as spatial filtration \citep{Gamarova2003,Rodriguez2004,Gamarova2005} or eclipse mapping \citep{Reed2001,Reed2005}. To further complicate this scenario, the highest amplitude pulsation mode observed in U~Gru ($f_{14}$) shows evidence for being tidally perturbed, adding an additional modulation to the observed pulsation amplitudes and phases.

We construct a toy model to illustrate the impact that eclipses have on the observed amplitudes and phases of non-tilted and tidally tilted pulsation modes\footnote{\url{https://github.com/colej/eb_mapping}}. The toy model is based on that used by Van~Reeth et al. (2022, {\it in prep.}), modified to our purposes. A full explanation of the code is given in Appendix~\ref{apdx:model}. In brief, this toy model assumes two spherical stars (with no deformation caused by the equilibrium tide) in a Keplerian orbit. As we are simply trying to illustrate the effect of time-dependent partial cancellation via eclipses and not attempting to forward model the effect, the use of spherical, non tidally-distorted stars is justified. The stars are inclined with respect to the observer according to the orbit. The pulsating component has its surface perturbed according to spherical harmonics of a given ($\ell$, $m$) combination. We then define a series of points in time, and calculate the total flux as observed considering the pulsation phase as well as the visibility of each star along the line of sight. This simple model allows us to track the observed pulsation phase and amplitude along the orbit as the pulsating star is eclipsed. Furthermore, we allow for the pulsation axis to be inclined such than it can align with the orbital plane to simulate the tidally tilted scenario, but we do not account for mode trapping. This toy model does account for linear limb-darkening, but does not account for the light travel time effect, which modulates the observed phase of a periodic pulsation because of the physical change in distance travelled by light from the pulsating star. In the case of U~Gru, this is a safe assumption as the peak-to-peak phase variation expected for $f_{14}$ caused by the light travel time effect is $\sim0.03$~rad, following \citet{Murphy2014}. 

\begin{figure}
\centering
\includegraphics[width=0.95\linewidth]{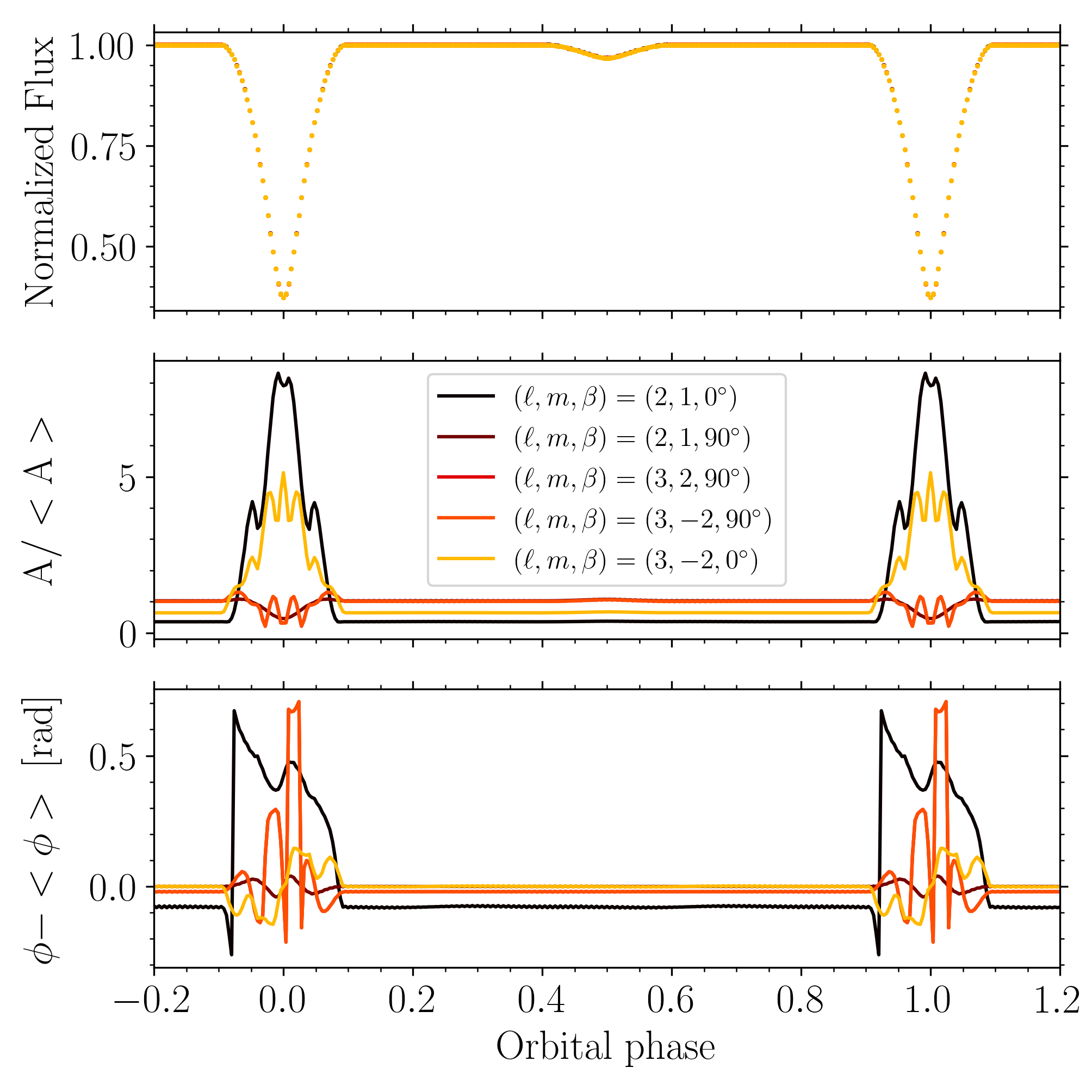}
    \caption{({\it Top}): Simulated light curves with different pulsation configurations ($\ell, m, \beta$) phase folded over the binary period. The points from the other simulated light curves are poorly visible due to the strong overlap between the models and the size of the markers. $(Middle)$: Mode amplitude variations for different pulsation configurations as seen by an observer because of changing partial cancellation effects. $(Bottom)$: Mode phase variations for different pulsation configurations as seen by an observer. }
\label{fig:geometric_model}
\end{figure}

We define the orbit and components of our toy model to be qualitatively similar to that of U~Gru following an initial analysis with the PHOEBE code \citep[PHOEBE 1; ][]{Prsa2005,Prsa2011}. This initial analysis uses the TESS light curve and {\sc uves} radial velocity solution (see Section~\ref{section:spectra}) and was adjusted by hand to match the light curve morphology and RVs of the primary. As the model was not numerically optimized, we do not consider this a full solution. We set the binary system to have $P_{\rm orb}=1.8805$~d, M$_1$=2.6~M$_{\rm \odot}$, R$_1$=2.8~R$_{\rm \odot}$, M$_2$=0.8~M$_{\rm \odot}$, R$_2$=2.6~R$_{\rm \odot}$, and $i$=85$^{\circ}$. We then simulate several pulsation configurations, each with a different mode geometry ($\ell,m$) and obliquity of the pulsation axis with respect to the rotation axis ($\beta$). We investigate multiple pulsation geometries in order to demonstrate that the differing pulsation geometries will produce different partial cancellation effects, resulting in different observed pulsation amplitudes and phases. The derived binary light curves, and pulsation amplitude and phase variations for each ($\ell, m, \beta)$ configuration are shown in the top, middle, and bottom panels of Fig.~\ref{fig:geometric_model}. The amplitude and phase modulations for the different pulsation configurations clearly demonstrate that the presence of eclipses can add an additional modulation to the  pulsation amplitude and phase at orbital phases when the pulsating component is eclipsed. Furthermore, the different models in Fig.~\ref{fig:geometric_model} demonstrate that the scale, complexity, and sign of the modulation are dependent on the system inclination, the obliquity of the pulsation axis, as well as the mode geometry \citep{Reed2005}.

Any additional periodic amplitude variability in a pulsation mode is captured in the Fourier transform of the time series. To investigate the impact that eclipse mapping has on the pulsation signal as seen in the periodogram, we simulate the light curve for two of the configurations shown in Fig.~\ref{fig:geometric_model} at the time stamps of the sector 1 TESS data. These simulated light curves are computed considering the full binary and pulsation signal, but have the eclipsing signature removed for clarity. By computing the synthetic light curves at the time stamps of the TESS data, we convolve the observational window function into the resulting periodogram to make our simulations as similar to the observations as possible. The periodograms of the subsequent time series are shown in Fig.~\ref{fig:eclipse_mapping_FT}. As it is a mathematical decomposition of a time series into a sum of sinusoids, the periodogram requires several additional sinusoidal terms to represent any non-sinusoidal variations, such as the amplitude modulations introduced by eclipse mapping. In this case, the signal manifests as multiple additional peaks, separated from the intrinsic pulsation frequency by integer multiples of the orbital frequency. 

The total number of additional peaks, their amplitudes, and whether or not they are symmetrically distributed about the intrinsic pulsation frequency depends on the mode geometry and orbital configuration. We highlight this by comparing the rows of Fig.~\ref{fig:eclipse_mapping_FT}. The top row considers an ($\ell=3, m=-2, \beta=0^{\circ}$) configuration where the two stars eclipse each other. In the configuration for the middle row, the radius of the secondary has been reduced so that the pulsating star is no longer eclipsed by the secondary, and therefore does not display any eclipse mapping effects. In this comparison, we note that the simulation without eclipses does not display the same series of peaks split by the orbital frequency, whereas the simulation with eclipses does. Furthermore, the bottom row displays a simulation for a different pulsation configuration ($\ell=2, m=-1, \beta=90^{\circ}$), again with the same orbital configuration as the top row that results in eclipses. This simulation demonstrates that the resulting signature in the periodogram is dependent on the pulsation geometry being considered, and that the resulting signature may have too low amplitude to be observed in real data with multiple noise components.

\subsection{Combined tidal and eclipse mapping effects}

\begin{figure*}
\centering
\includegraphics[width=0.95\linewidth]{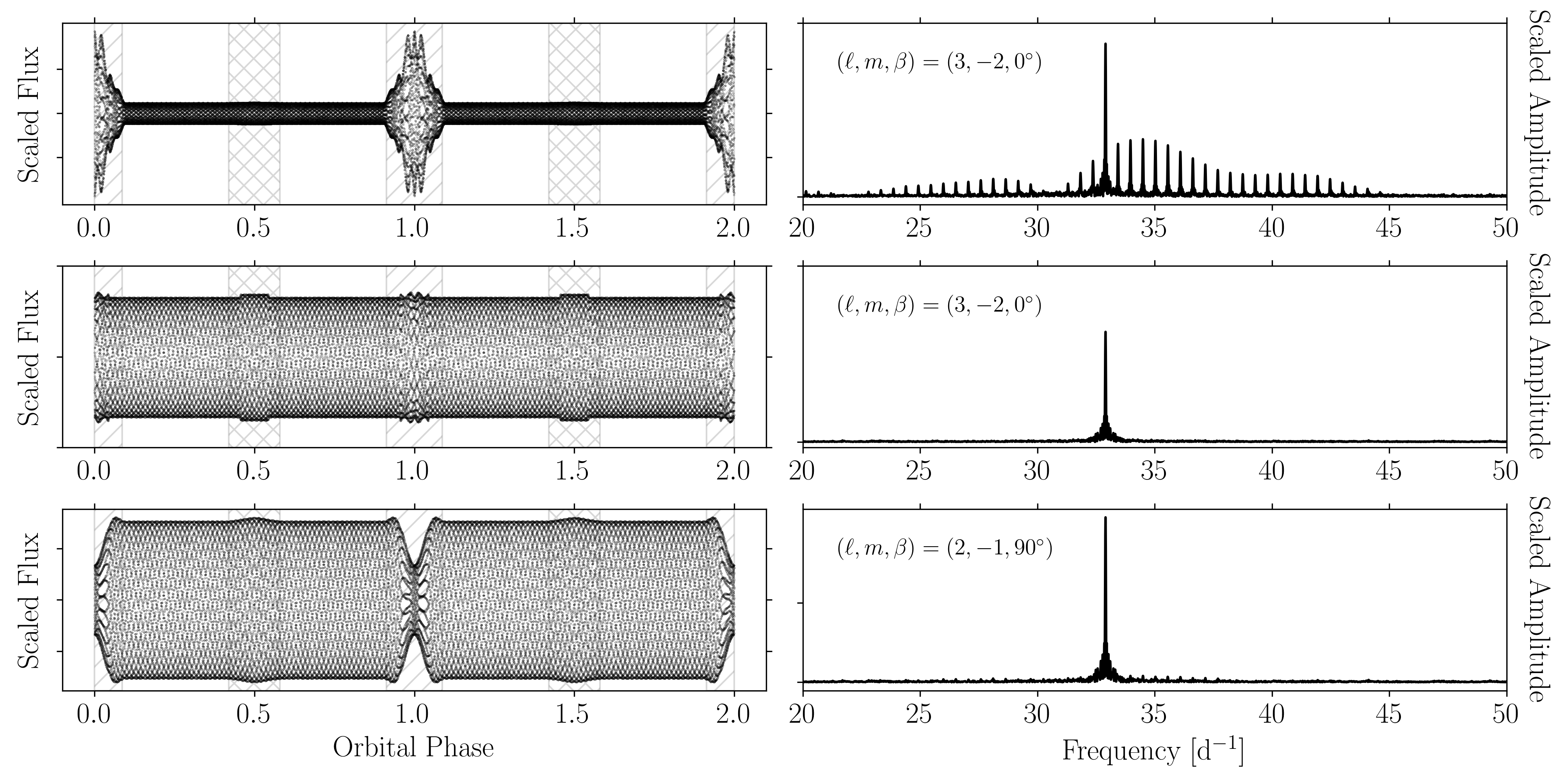}
    \caption{Left column display simulated light curves of the pulsation mode only. Right column displays periodogram of the accompanying light curve. These light curves do not consider the light variations caused by the binary itself. Grey hatching denotes phases of primary (forward hatching) and secondary (cross hatching) eclipse.}
\label{fig:eclipse_mapping_FT}
\end{figure*}

Our toy model produces two quantifiable outputs: 1) complex behaviour in the tracked mode amplitude and phase that occurs specifically during eclipse of the pulsating component, and 2) a series of peaks in the periodogram that are separated by the orbital frequency and potentially asymmetrically distributed around the true pulsation frequency. The observed chaotic mode amplitude and phase behaviour during phases of primary eclipse (e.g. Fig.~\ref{fig:obs_modulations}) is qualitatively similar to the behaviour predicted by our eclipse mapping model. Given this, we propose that the asymmetrically distributed series of peaks separated by the orbital frequency observed in the residual periodogram of U~Gru is also caused by eclipse mapping, similar to the results in Fig.~\ref{fig:eclipse_mapping_FT}. Following this interpretation, we can use the mathematical framework presented by \citet{Jayaraman2022} to investigate the observed amplitude and phase of this mode across the orbit and during primary eclipse. This framework assumes that any frequencies separated from the intrinsic mode by a multiple of the orbital frequency can be used to derive information about the phase and amplitude modulation of the intrinsic pulsation mode. While this methodology was developed to interpret the case of tidally tilted pulsators, it can be generalised for other pulsations that show orbital phase dependence. We simultaneously fit for the amplitude, and phase of the mode at $f_{14}=33.046~$d$^{-1}$, as well as the amplitudes and phases of those frequencies located at integer multiples of the orbital frequency such $f_j = f_1 \pm n_j$ with $n$ being a real integer. We fit for the 14 frequencies with $f {\rm mod} f_{\rm orb}<0.1$~d$^{-1}$  simultaneously, all of which are taken from Table~\ref{tab:ipw_results}. In addition to the frequency at $f_{14}=33.046~$d$^{-1}$, we fit 12 sinusoids that are at frequencies below the main frequency, and only one at higher frequencies. While there may be other components to this series of frequencies, they do not have high enough amplitudes to be considered significant in our frequency analysis, and are therefore ignored. Although \citealt{Handler2020,Kurtz2020, Rappaport2021,Jayaraman2022} intentionally set the reference epoch for the pulsations to occur when the $f_j=n\pm$1 frequencies have maximum amplitude, we cannot follow the same procedure as the frequencies we consider are asymmetrically distributed in the case of U~Gru. Instead, we retain the same reference epoch as that of superior conjunction reported in \citet{Bowman2019}. 

\begin{figure}
\centering
\includegraphics[width=0.95\linewidth]{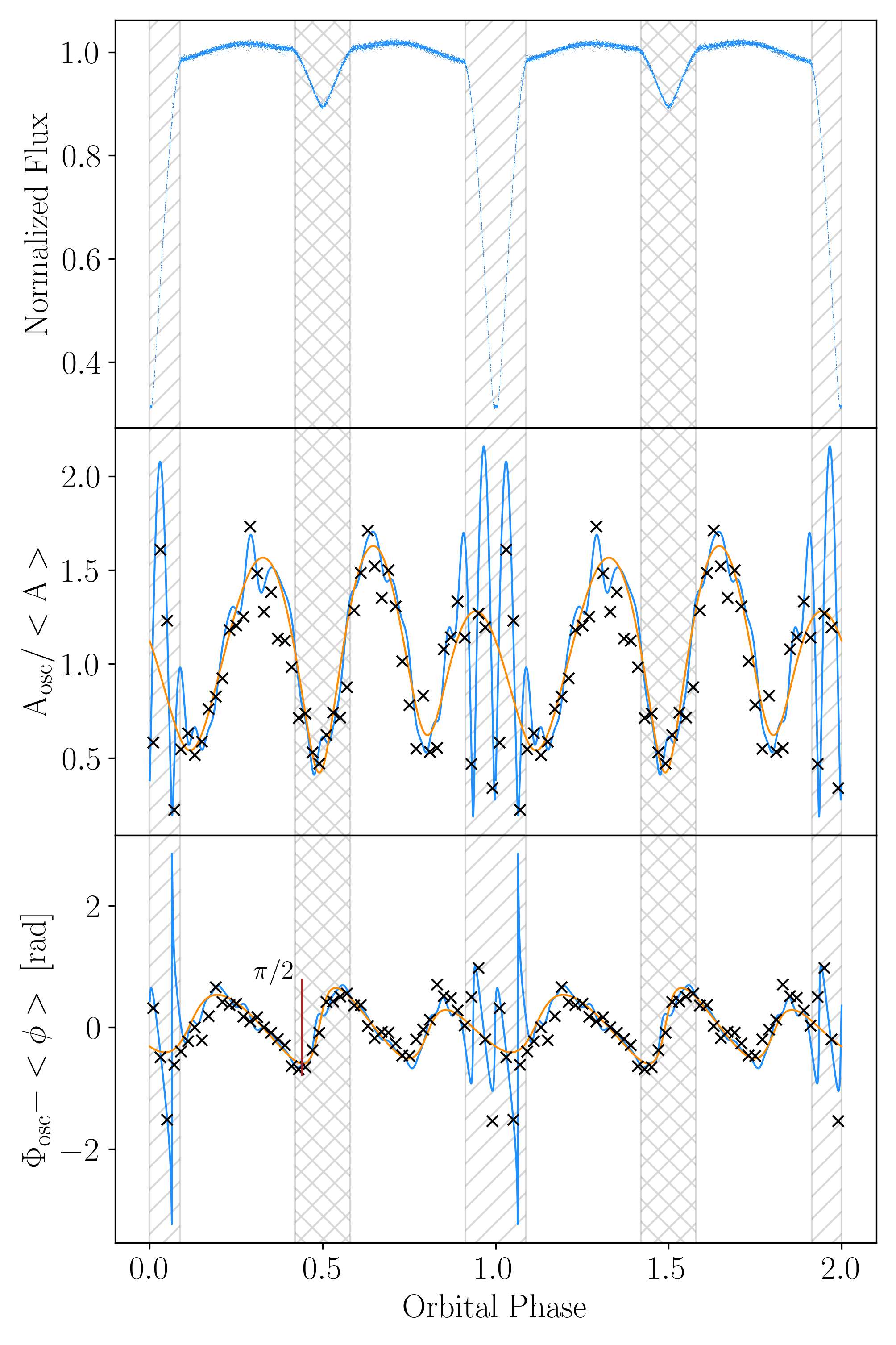}
    \caption{{\it (Top)}: Phase folded light curve of U~Gru for sector 1. {\it (Middle)}: Amplitude modulation of $f_{14}=33.046$~d$^{-1}$ as reconstructed according to the fitting method developed by \citet{Jayaraman2022}. The blue considers all 14 significant frequencies split from $f_{14}$ listed in Table~\ref{tab:ipw_results}. The orange lines only consider those frequencies with $f=f_{14}\pm Nf_{\rm orb}$, where $-3 \le N \le 3$, if they were found to be significant in the iterative pre-whitening procedure. Tracked mode amplitudes from Section~\ref{section:mode_tracking} shown as black x's. {\it (Bottom)}: Same as the middle panel but for the phase modulation. Grey hatching denotes phases of primary (forward hatching) and secondary (cross hatching) eclipse.}
\label{fig:phamp}
\end{figure}

The resulting amplitude and phase modulation over the orbit are shown in the middle and bottom panels of Fig.~\ref{fig:phamp}. The blue trend represent the fits using sector 1 data. The general phase and amplitude behaviour recovered by this method agrees with the behaviour recovered by fitting a sinusoid with $f=33.0456~$d$^{-1}$ to different phase bins carried out in Section~\ref{section:tidal_effects} (shown in black in Fig.~\ref{fig:phamp}). We also find complex amplitude and phase modulations around primary eclipse in Fig.~\ref{fig:phamp}. We further construct a model considering only those frequencies separated from the intrinsic mode by $-3\le N\le3$ (if significant). This model is shown in orange in the middle and bottom panels of Fig.~\ref{fig:phamp}, respectively. When only considering frequencies separated from the intrinsic mode by $\pm3$ times the orbital frequency, we only recover the three amplitude maxima and phase sign changes, and not the complex behaviour near primary eclipse. This suggests that the remaining asymmetrically distributed frequencies separated by larger integer multiples of the orbital frequency are the result of the complex behaviour at primary eclipse. We argue that this further supports the eclipse mapping origin of the asymmetrically distributed series of frequencies separated by the orbital frequency. We find the same results when using the sector 28 data.

\section{Discussion and conclusions}
\label{section:conclusions}

To date there are multiple examples of pulsating stars in binaries that exhibit behaviour consistent with the tidally tilted or trapped pulsator model. In addition to this, there are an increasing number of systems that exhibit pulsational behaviour that cannot be explained by the existing tidally tilted or trapped models. In this work we increase the diversity of observed behaviour with the case of the bright triple system U~Gru. 

We used {\sc uves} spectroscopy to confirm the Algol-type status of U~Gru and further discover that the system consists of an apparently rejuvenated A-type star and cool companion in a tight 1.8805~d orbit with a wide F3V tertiary companion. The atmospheric solution for the primary is consistent with a moderately rotating, early A-type star that has been rejuvenated by recent mass transfer. Unfortunately, due to its low light contribution, we could not robustly detect the signature of the cool secondary component. The lack of RV shifts and eclipse timing variations in the TESS data suggests that the orbit of the tertiary is much longer than the two year time-base covered by the TESS data. To this end, we consider the separation sufficiently large as to not introduce additional tidal forcing from the distant third body. 

The first frequency analysis of the TESS light curve by \citet{Bowman2019} revealed a series of significant frequencies separated by the orbital frequency, or integer multiples thereof. In this work, we conducted a follow-up analysis of the TESS light curves for both sectors 1 and 28. We removed the dominant binary signal via an interpolated model and ran an iterative pre-whitening frequency analysis on the residual light curves, recovering the independent pulsations as well as the series of frequencies split by integer multiples of the orbital frequency. This demonstrated that the mechanism which produced the series of frequencies separated by multiples of the orbital frequencies is stable over the 2-yr TESS time base. 

The multiperiodic pulsations of U~Gru present a challenge for interpretation from only a single tidal mechanism. Only one pulsation mode ($f_{14} = 33.046~$d$^{-1}$) shows evidence for amplitude and phase modulation across the binary orbit similar to what is expected in the case of a tidally tilted pulsation axis. However, the odd number of amplitude minima and maxima and phase zero-crossings, combined with the $\pi/2$~rad phase variation are not consistent with predictions from the tidally tilted or trapped pulsator models \citep{Fuller2020}. Though, it is possible that future developments of this model that include the Coriolis force could produce predictions that more closely match our observations. 

Currently, the $\pi/2$~rad phase variations we find are most similar to that of V456~Cyg, where small phase and normal amplitude variations were observed in a subset of g modes in a circular, synchronised binary \citep{VanReeth2022}. Because of this, the authors argue that the modes are tidally perturbed and have contributions from non-axisymmetric components. Despite the similarities, it is not clear whether or not the same mechanism that modifies the g modes in V456~Cyg can modify the p modes in U~Gru. Alternatively, due to the moderate rotation of the pulsating A-star, we could be observing the effects of rotation modifying and confining p modes to `island modes' \citep[e.g., ][]{Ouazzani2012,Reese2013,Reese2017}. Moderate to rapid stellar rotation can preferentially trap pulsations to specific latitudes and longitudes, depending on the mode geometry and rotation rate \citep{Aerts2010}. As such, moderate rotation that is synchronised with the binary orbit could, in theory, produce an observed amplitude modulation that is commensurate with the orbital period. However, there is no prediction for the observed phase behaviour of such rotationally modified modes. It is also worth noting that while U~Gru hosts multiple independent pulsation modes, most do not show any amplitude or phase variation with the orbit. This suggests that the mechanism behind the observed amplitude and phase variation in the dominant pulsation mode is not ubiquitous for all modes. Interestingly, other systems such as RS~Cha and HD~265435 also host some pulsations that display amplitude and phase modulation and some others that do not \citep{Steindl2021,Jayaraman2022}. 

The observed series of frequencies that are separated by (multiples of) the orbital frequency cannot be explained by any currently proposed tidal mechanism. Instead, we present a toy model to demonstrate that the observed signals can be explained by eclipse mapping. The results of the toy model demonstrate that eclipse mapping can produce a series of frequencies that are separated by the orbital frequency and that are asymmetrically distributed around the intrinsic pulsation frequency, as is observed in the case of U~Gru. This model also demonstrates that the specific resulting signature in the periodogram depends on the mode geometry, such that some mode geometries might not experience any observable effect from eclipse mapping for a given orbital configuration. The mode geometry dependence can explain why we observe some pulsation modes in U~Gru to experience eclipse mapping effects, while others to not although they originate from the same system. This logic can be further extended to other pulsating stars in eclipsing binaries such as RS~Cha, TIC~63328020, TZ~Dra, HL~Dra, V1031~Ori, and VV~Ori, where some show complex series of frequencies separated by the orbital frequency and others do not  \citep{Steindl2021,Rappaport2021,Lee2021,Southworth2021,Shi2021,Kahraman2022}. 

While it has been hypothesised that eclipse mapping can serve as a means of mode identification for pulsations in well characterised eclipsing binary systems, there are numerous observational considerations to account for \citep{Rodriguez2004,Reed2005,Gamarova2005}. Primarily, the amplitude of the eclipse mapping effect depends on the mode geometry and amplitude, as well as the specifics of the binary orbit. To this end, unambiguous mode identification through eclipse mapping would require the observer to be able to identify all of the frequencies distributed about a given pulsation frequency. Given the typical instrumental and astrophysical noise components in astronomical time series, this task is challenging. Furthermore, if any additional tidal phenomena are active which produce frequencies separated from the pulsation frequency by the orbital frequency, a comparison to theoretical predictions would need to account for their contributions as well. Although, compared to the model dependent methods of mode identification via amplitude ratios or phase differences from multi-colour photometry \citep{Garrido2000,Handler2008}, eclipse mapping provides an alternative that does not depend on the underlying detailed stellar model.

U~Gru's status as an oEA star places it in a growing subset of stars with some form of tidally influenced pulsations that have a recent history of mass transfer. Other oEA stars in this category include TZ~Dra, HL~Dra, V1031~Ori, and V456~Cyg. Future studies that focus on pulsations in oscillating Algol type systems would benefit from investigating for the signature of tidally tilted pulsation axes, or otherwise tidally influenced pulsations in a systematic manner. As we are observing the whole sky with high precision space-based data, we expect to find more systems with an increasing diversity of tidally related phenomena. Increasing the sample of such pulsators will enable us to investigate the underlying physical mechanisms, as well as investigate their dependence on stellar properties and potentially their evolutionary history. Future missions such as  PLATO \citep{Rauer2014} will be of great benefit in this realm as it will deliver multi-colour photometry. With high-cadence multi-colour photometry, future work will be able to independently constrain mode identification to further scrutinise the underlying mechanisms behind the increasing diversity of observed tidally related phenomena. Indeed, future work should investigate the wavelength dependence to the observed effects of eclipse mapping, and the extent of the combined power of mode identification through the combination of eclipse mapping and mulit-filter amplitude and phase ratios, which often suffer from limitations when individually applied \citep{Daszynska2002,Dupret2003}.

\section*{Acknowledgements}
We thank the anonymous referee for their comments which improved this manuscript. CJ acknowledges support from the Netherlands Research School of Astronomy (NOVA). The research leading to these results has received funding from the European Research Council (ERC) under the European Union's Horizon 2020 research and innovation programme (grant agreement N$^\circ$670519: MAMSIE; CJ, DMB, AT), from the Research Foundation Flanders (FWO) under grant agreement G0A2917N (BlackGEM; CJ), from the KU\,Leuven Research Council (grant C16/18/005: PARADISE; AT), from the Research Foundation Flanders (FWO) under grant agreements G0H5416N (ERC Runner Up Project; AT), as well as from the BELgian federal Science Policy Office (BELSPO) through PRODEX grant PLATO (AT). TVR and DMB gratefully acknowledge financial support from FWO under grant agreement numbers 12ZB620N and 1286521N, respectively.

This work is based on observations collected at the European Southern Observatory under ESO programme 0104.D-0209. This paper includes data collected by the TESS mission. Funding for the TESS mission is provided by the NASA's Science Mission Directorate. This paper includes data collected by the TESS mission, which are publicly available from the Mikulski Archive for Space Telescopes (MAST). This work has made use of the VALD database, operated at Uppsala University, the Institute of Astronomy RAS in Moscow, and the University of Vienna.

This work has made use of the {\sc numpy} \citep{Harris2020}, {\sc matplotlib} \citep{Hunter2007}, {\sc pymc3} \citep{salvatier2016}, and {\sc exoplanet} \citep{ForemanMackey2021} python packages.

\bibliographystyle{aa}
\bibliography{u_gru_II.bib}

\newpage

\appendix
\section{Pulsating binary model} 
\label{apdx:model}

We construct a binary system consisting of two spherical, non-tidally distorted stars in a Keplerian orbit. The orbit is defined by the masses of the two stars (M$_{1,2}$), their orbital period (P$_{\rm orb}$), reference epoch of superior conjunction ($t_{0,sc})$ , and separation ($a$), and the inclination of their orbit with respect to the observers line of sight ($i$). In addition to their masses, the stars are characterised by their radii (R$_{1,2}$) and what percent of the total light seen by the observer they contribute ($\ell_{1,2}$). Furthermore, we include a third light contribution that has ambiguous origin.

The surface of each star is discretized on a spherical grid with radius r=R$_{\star}$, latitude $-\pi \le \theta \le \pi$, and longitude $0\le \phi \le 2\pi$. At any point in time, only a particular range of latitudes and longitudes on the stellar surface are visible, denoted as ($\theta_0, \phi_0$)$_t$. The user is allowed to decide if either or both stars are pulsating, and assign a mode identification ($\ell, m$) and frequency ($f_{\rm pulse})$ to the pulsation mode. For this model, only one pulsation per star is allowed. The system, and therefore the surface of each star is inclined according to the binary inclination, changing the visible range of latitudes and longitudes on the stellar surface. The inclined surface ranges are denoted as ($\theta_i, \phi_i$)$_t$. Finally, we allow for the surface to be inclined again by some angle $\beta$ to simulate the case of a tidally tilted pulsation. The final visible surface is then denoted by ($\theta_{\beta}, \phi_{\beta}$)$_t$. We then calculate the perturbation to the surface brightness following the given mode geometry according to:
\begin{equation}
    \begin{gathered}
        L_r = a_r Y_{\ell}^m \left( \theta, \phi \right) \exp \left( -i 2\pi f t\right) \\ 
        L_{\theta} = b_r \frac{ \partial Y_{\ell}^m \left( \theta, \phi \right)}{\partial \theta} \exp \left( -i 2\pi f t\right) \\
        L_{\theta} = \frac{b_r}{\sin\theta} \frac{ \partial Y_{\ell}^m \left( \theta, \phi \right)}{\partial \phi} \exp \left( -i 2\pi f t\right), \\
    \end{gathered}
\end{equation}
where $Y_{\ell}^m \left( \theta, \phi \right)$ are spherical harmonics, and $a_r$ and $b_r$ are the radial and horizontal amplitudes at the stellar surface, related through the pulsation frequency in the corotating frame $f_{co}$ by:
\begin{equation}
    \label{eqn:arbr}
    \frac{a_r}{b_r} = \frac{G M_{\star}}{f_{co}^2 R_{\star}^3}.
\end{equation}
For simplicity, we set $a_r=1$ and calculate $b_r$ through Eq.~\ref{eqn:arbr}.

We initialise a grid of time points, and progress the positions of the stars according to the orbit. At each time point, we check if the surface of either star is being eclipsed, and if so apply a simple boolean mask to the points on the stellar surface that are in eclipse. We calculate the summed fluxes of the stars according to their visible (and potentially pulsating) surfaces and scale them according to their light contributions considering any potential third light. We provide the total summed binary and pulsation flux, as well as the individual fluxes of each component.

\label{lastpage}

\end{document}